\documentclass[twocolumn,aps,floatfix,showpacs,superscriptaddress]{revtex4}
\usepackage{amsmath,amssymb,eucal,graphicx,float,epstopdf}

\usepackage{ulem}
\usepackage{color}

\begin{document}
\title{Generalized exclusion processes: Transport coefficients}

\author{Chikashi Arita}
\affiliation{Theoretische Physik, Universit\"at
 des Saarlandes, 66041 Saarbr\"ucken, Germany}
\email{c.arita@physik.uni-saarland.de}
\author{P. L. Krapivsky}
\affiliation{Department of Physics, Boston University, Boston, MA 02215, USA}
\email{pkrapivsky@gmail.com}
\author{Kirone Mallick}
\affiliation{Institut de Physique Th\'eorique, IPhT, CEA Saclay
and URA 2306, CNRS, 91191 Gif-sur-Yvette cedex, France}
\email{kirone.mallick@cea.fr}


\begin{abstract}
 A class of generalized exclusion processes with symmetric nearest-neighbor hopping which are  parametrized by the maximal occupancy, $k\geq 1$, is investigated. For these processes on hyper-cubic lattices we compute the diffusion coefficient in all  spatial dimensions. In the extreme cases of $k=1$ (symmetric simple exclusion process) and $k=\infty$ (non-interacting symmetric random walks) the diffusion coefficient is constant, while for $2\leq k<\infty$ it depends on the density and $k$. We also study the evolution of the tagged particle, show that it exhibits a normal diffusive behavior in all dimensions, and probe numerically the coefficient of self-diffusion. 
\end{abstract}
\maketitle

\section{Introduction}

Exclusion processes constitute an important class of lattice gases
that plays a prominent role in numerous subjects including non-equilibrium
statistical mechanics, soft matter, traffic models, biophysics, 
combinatorics and  probability theory \cite{KLS,Spohn,SZ95,D98,L99,KL99,KRB10,Schad,CKZ}.
By definition, exclusion processes are lattice
gases supplemented with stochastic hopping and obeying the constraint
that at most one particle per site is allowed. In simple exclusion models,
only hops to nearest-neighboring sites are allowed.  These models are
exactly solvable in low dimensions \cite{D98,S00,BE07}, and they have become
benchmarks to test general theories for  non-equilibrium behaviors
\cite{DerrReview,Bertini,Bodineau,DCairns,JonaSeoul,MFTReview}.

\begin{figure}
\centerline{ \includegraphics[width=8.7cm]{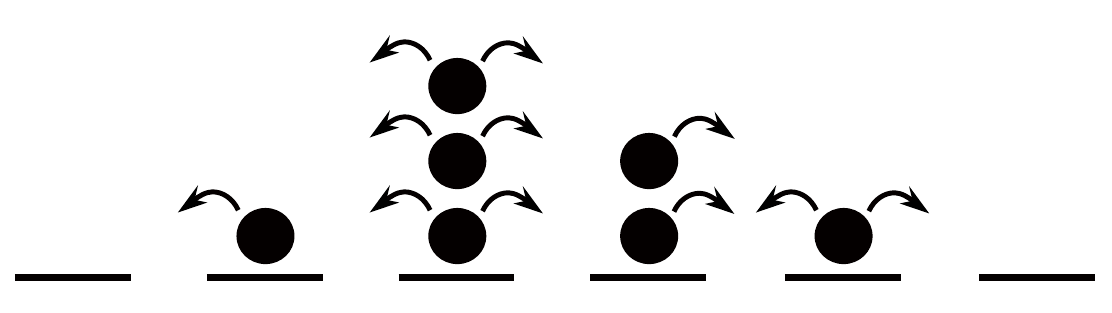}  }
\vspace{-0.23cm}
\caption{GEP with $k=3$ in one dimension. Hopping into a site occupied by three particles is forbidden; other hopping events occur with the same (unit) rate. Thus the total hopping rate from every site is equal to the number of particles at the site times the number of neighboring sites which are not fully occupied. }
\label{pic_fig}
\end{figure}

Because of its ubiquity and usefulness, numerous more complicated variants of the 
basic exclusion process have been investigated
(see \cite{SZ95,S00,CKZ} and references therein). One natural generalization
is to alleviate  the exclusion constraint by allowing $k$ particles per site ($k \ge 1$ is a fixed integer).
More precisely, this process is defined  in $d$ dimensions, say on the hyper-cubic
lattice $\mathbb{Z}^d$, as follows: (i) each  particle attempts to hop
to its $2d$ neighbors with the same unit rate to each
neighbor (symmetric hopping), (ii) every hopping attempt is successful when the target site is
occupied by less than $k$ particles, otherwise the hopping attempt is
rejected  (Fig.~\ref{pic_fig}). The symmetric exclusion process (SEP) is recovered when $k=1$, whereas for $k=\infty$ the model reduces to independent random walks undergoing symmetric nearest-neighbor hopping. Letting $k$ vary from 1 to $\infty$ allows us to interpolate between a strongly interacting to a non-interacting system.  (In quantum systems, restricting the maximum number of particles in a given quantum state to an integer $k$,
with $1 < k < \infty$, leads to the so-called Gentile statistics \cite{Gentile, Khare} interpolating between Fermi-Dirac statistics  and Bose-Einstein statistics. In this paper we consider only classical lattice gases.)

Generalized exclusion processes (GEPs) with maximal occupation number $1<k<\infty$ have been studied in \cite{KLO94,Kipnis2,Timo}; see also 
Refs.~\cite{SandowSchutz,Barma_drop,Barma,Schutz,PK,Kurchan,CM,CGRS,Ryabov,Becker_13,Becker_preprint}
for other versions of GEPs. Some of these models  can be mapped onto multi-species exclusion processes \cite{MMR,EFM,FerrariMartin} but with non-conserving species. Overall, the GEPs are considerably less understood than the ordinary exclusion process---the integrability properties of the SEP do not carry over to the GEPs and a different perspective is required. 

The objective of  this work is to investigate the GEP at a coarse-grained level and to calculate the transport coefficients that underlie the hydrodynamic description. Section \ref{sec:HB} reviews known  properties of the GEP and outlines its macroscopic (i.e., hydrodynamic) regime which is governed by a diffusion equation. We also formulate our main result, namely a parametric representation of the diffusion coefficient. In Sec.~\ref{sec:DC},  we present the derivation of the diffusion coefficient. In Sec.~\ref{sec:SDP}, stationary density profiles are computed and compared with simulations.
In Sec.~\ref{sec:SDC},  we examine the evolution of a tagged particle in the GEP in various spatial dimensions. We derive the self-diffusion coefficient using a mean-field approximation. We also probe the self-diffusion coefficient numerically and show a reasonable 
qualitative agreement with mean-field predictions.  We summarize our results, and mention remaining challenges and possible extensions, in Sec.~\ref{sec:Sum}.

\section{Hydrodynamic Behavior}
\label{sec:HB}

For the generalized exclusion process with symmetric hopping, steady states are remarkably
simple and are given by a product measure \cite{Spohn,L99,KL99,PK}.  In other words, one only needs to know the probabilities $P_j$ to have $j$ particles per site and  the stationary weight of any configuration factors  into the product of these basic probabilities. The basic probabilities are given by an elementary  formula \cite{Barma,KL99,PK}
\begin{equation}
\label{probs}
 P_j = \frac{\lambda^j}{j!}\,\frac{1}{Z_k(\lambda)}\, .
\end{equation}
To justify \eqref{probs} it suffices to use the factorization and verify that the flow $(i,j)\Longrightarrow (i-1,j+1)$, 
which is given by $iP_iP_j$, is equal to the flow $(i-1,j+1)\Longrightarrow (i,j)$, 
which is given
by $(j+1)P_{i-1}P_{j+1}$. With the choice \eqref{probs},  we indeed get
$iP_iP_j = (j+1)P_{i-1}P_{j+1}$. The  `partition function'  $Z_k(\lambda)$, which appears in
Eq.~\eqref{probs}, is fixed by the normalization requirement
$\sum_{0\leq j\leq k}P_j=1$. It is equal to an incomplete exponential function:
\begin{equation}
\label{partition}
  Z_k(\lambda) = \sum_{j=0}^k \frac{\lambda^j}{j!}\, .
\end{equation}
The `fugacity' parameter $\lambda$ is implicitly determined by the
density $\rho$:
\begin{equation}
\label{density}
\rho = \sum_{j=0}^k jP_j = \lambda\,\frac{Z_{k-1}(\lambda)}{Z_k(\lambda)}\,.
\end{equation}
Furthermore, GEPs with rather general hopping rates depending on the number of particles on the exit site have been also studied; see, e.g. \cite{KL99,KLO94,Barma_drop,Barma,PK,YN_06}. In all these models the steady-state probabilities are also given by a product measure. We emphasize that this product measure structure is akin to that of the zero-range process \cite{MartinZRP,DeMasi} although the two processes are fundamentally different (in the GEP, the jump rate of a particle does depend on the state of the target site, contrary 
to what is assumed in the zero-range process).

In order to study the dynamics of the system, the knowledge of the
steady-state distribution is not sufficient and a full description
of the evolution requires the complete spectrum and eigenstates
of the evolution matrix. Yet the large scale ``hydrodynamic''  behavior is conceptually simple. The
only relevant hydrodynamic variable is density and it evolves
according to a diffusion equation. In one dimension, for instance, it reads 
\begin{equation}
\label{DE}
\frac{\partial \rho}{\partial t} = \frac{\partial}{\partial x}
\!\left[D_k(\rho)\frac{\partial \rho}{\partial x}\right] .
\end{equation}
This generic result is valid for lattice gases with symmetric hopping
\cite{Spohn,KL99,L99}. Thus the detailed microscopic rules underlying the dynamics of the lattice gas play little role; namely, they are all encapsulated in a single density-dependent function, the diffusion coefficient. 

The determination of the diffusion coefficient is in principle a very difficult problem as we do not assume the
lattice gas to be dilute. For the GEP, the diffusion coefficient $D_k$ is known
in the extreme cases, namely for symmetric random walks ($k=\infty$)
and for the SEP ($k=1$). In both these cases the diffusion coefficient
is constant; with our choice of the hopping rates, we have 
\begin{equation}
\label{Diff}
D_1 = D_\infty = 1.
\end{equation}
For other maximal occupancies ($1<k<\infty$), the diffusion
coefficient is density-dependent. This already follows from the
asymptotic behaviors 
\begin{equation}
\label{Diff_asymp}
D_k(\rho) = 
\begin{cases}
1 & \rho\to 0 ,\\ k & \rho\to k .
\end{cases}
\end{equation}
The small-density asymptotic corresponds to the diffusion of a single
particle in the empty system, while the behavior in the $\rho\to k$
limit can be understood by considering a single vacancy in the fully occupied system. 

\begin{figure}
\centerline{ \includegraphics[width=88mm]{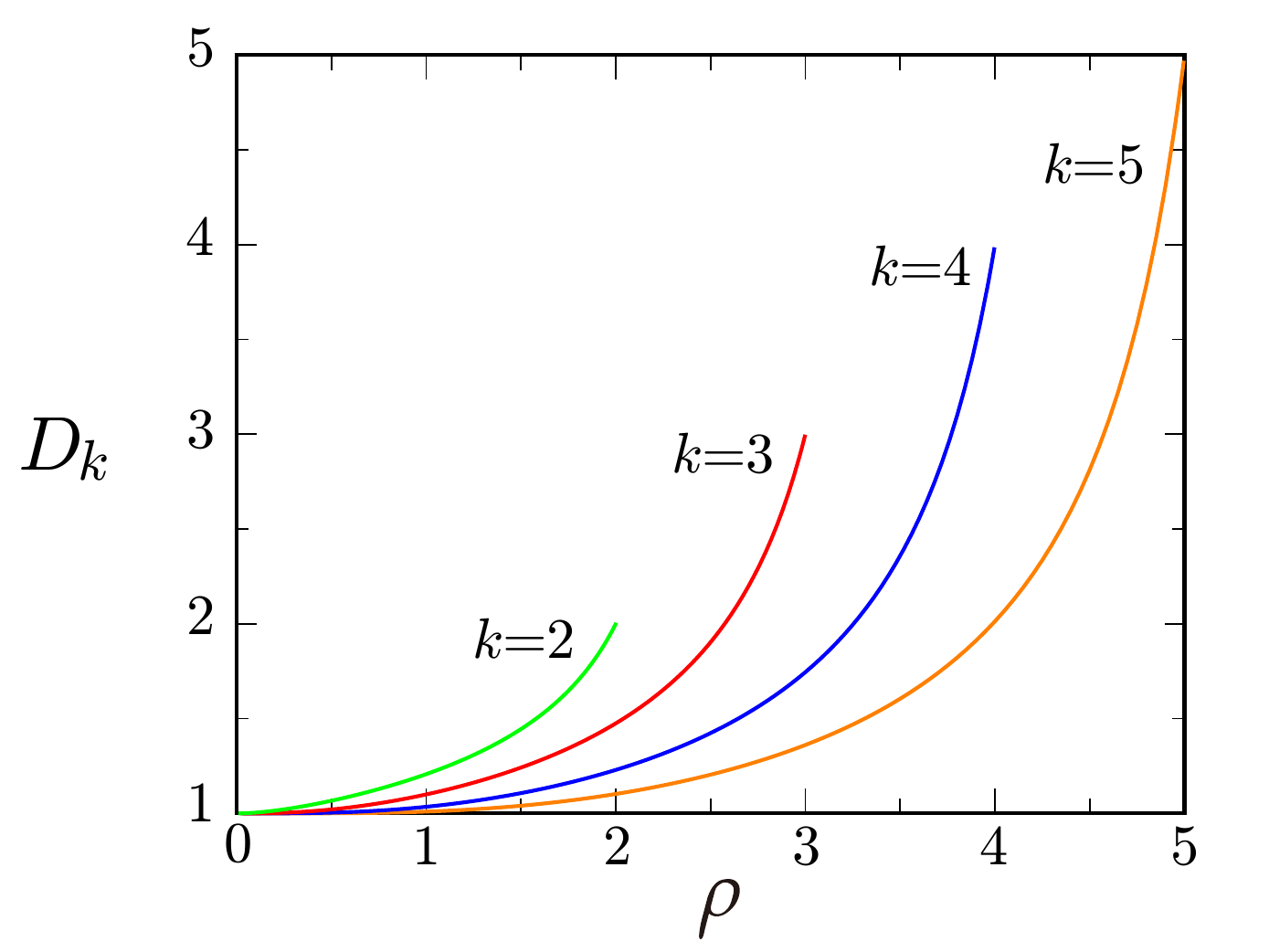}  }
\caption{Diffusion coefficient \eqref{Dk} as a function of density for
  the GEP with $k=2,3,4,5$.}
\label{D5_fig}
\end{figure}

The computation of $D_k(\rho)$ for all $k$ will be presented
in  section~\ref{sec:DC}.  We will show that 
\begin{equation}
\label{Dk}
D_k = \Lambda_k - \rho\,\frac{d\Lambda_k}{d\rho}\,, \quad
\Lambda_k(\rho)=1-P_k(\rho) .
\end{equation}
Using \eqref{probs} and \eqref{density} one obtains the parametric
representation of $\Lambda_k(\rho)$:
\begin{equation}
\label{P_max}
\rho =  \lambda\,\Lambda_k(\lambda)\,,\quad \Lambda_k(\lambda) =
\frac{ Z_{k-1}(\lambda)}{ Z_k(\lambda)}  .
\end{equation}
These formulas apply to all $1\leq k\leq \infty$ including the extreme
cases. For the SEP we have $\Lambda_1=1-\rho$, while for random walks
$\Lambda_\infty=1$; in both cases we recover \eqref{Diff}. In all other cases ($2\leq k<\infty$), 
the diffusion coefficients $D_k(\rho)$ are monotonically increasing convex functions of the density 
(as illustrated in Fig.~\ref{D5_fig} for $k=2,3,4,5$).

For lattice gases in higher dimensions, the density generally
satisfies a diffusion equation
\begin{equation}
\label{rho:eq}
\frac{\partial \rho}{\partial t} = \sum_{a,b=1}^d \frac{\partial
}{\partial x_a} \left[D^{ab}(\rho)  \frac{\partial  \rho}{\partial
    x_b}\right]  
\end{equation}
with a $d\times d$ diffusion matrix $D^{ab}(\rho)$. An ordinary diffusion process (e.g., a symmetric random walk) is
macroscopically isotropic, so the diffusion matrix is scalar: $D^{ab}(\rho)=\delta^{ab}D(\rho)$.
Generally the diffusion matrix is symmetric, $D^{ab}(\rho)=D^{ba}(\rho)$, and for lattice gases on $\mathbb{Z}^d$ the symmetry of the lattice limits the number of independent matrix elements to two: All diagonal elements are equal [we denote them by $D(\rho)$], and all off-diagonal are also equal  [we denote them by $\widehat{D}(\rho)$]. In three dimensions, for instance, the diffusion matrix is
\begin{equation}
\label{matrix}
{\bf D}(\rho) = \left[ \begin{array}{ccc} D(\rho)
  &\widehat{D}(\rho) & \widehat{D}(\rho)\\ \widehat{D}(\rho) & D(\rho)
  & \widehat{D}(\rho)\\ \widehat{D}(\rho) & \widehat{D}(\rho) &
  D(\rho)
\end{array} \right].
\end{equation}

For lattice gases on $\mathbb{Z}^d$ in which each particle occupies a single lattice site one expects the diffusion matrix to be scalar and the diffusion coefficient  to be strictly positive \cite{critical}. As far as we know these physically obvious assertions have not been proved in full generality. In the next section, we calculate $D_k(\rho)$ for all $k$. Our derivation implies that for the GEP the diffusion process is indeed macroscopically isotropic, and our explicit results show that the diffusion coefficients satisfy $D_k(\rho)>1$ and they are independent of the spatial dimension.

\section{Diffusion Coefficient}
\label{sec:DC}

In this section, we calculate the diffusion coefficient for the GEP that appears in 
Eq.~(\ref{rho:eq}). As a warm-up, we recall the well-known case of the SEP. Then we show the crucial difference between the GEPs with $2\leq k<\infty$ and the SEP, namely the presence of the higher-order correlators, which makes impossible an elementary derivation of the diffusion coefficient. Fortunately, the understanding of the equilibrium in the GEP and a perturbative expansion around the equilibrium gives a method for deriving the diffusion coefficient for the GEPs with $2\leq k<\infty$.  We present a detailed derivation of $D_2(\rho)$ in one and higher dimensions. The case of arbitrary $k$ is outlined at the end of this section.

A configuration of the SEP on a one-dimensional lattice is fully described by binary variables
$n_j(t)$: If the site $j\in\mathbb{Z}$ is empty, $n_j(t)=0$; if it is
occupied,  $n_j(t)=1$. In an infinitesimal time interval $dt$, the
particle hops from site $j$ to site $j+1$ with probability
$n_j(1-n_{j+1})dt$. This choice assures that the hopping event happens
only when the site $j$ is occupied and the site $j+1$ is empty. Taking
into account all possible hops one finds that the average density
evolves according to
\begin{align}
\begin{split}
\frac{d \langle n_j\rangle}{dt} &=  \langle n_{j-1}(1-n_j)
+n_{j+1}(1-n_j)\rangle\\ &- \langle
n_j(1-n_{j-1})+n_j(1-n_{j+1})\rangle \, ,
\end{split}
\end{align}
which simplifies to the discrete diffusion equation
\begin{equation}
\label{DDE}
\frac{d \langle n_j\rangle}{dt} = \langle n_{j-1}\rangle - 2\langle
n_j\rangle + \langle n_{j+1}\rangle \, .
\end{equation}
The remarkable cancellation of the higher-order correlation functions
allows one to prove the validity of the hydrodynamic limit without
further assumptions---no need to use the absence of correlations in
the steady state. By definition, in the hydrodynamic limit the average
density varies on the scales greatly exceeding the lattice
spacing. Therefore we write $\langle n_j(t)\rangle =\rho(x,t)$; the
notation $x=j$ emphasizes that we are switching to the continuum
description. We then expand $\langle n_{j\pm 1}\rangle$ in Taylor series 
\begin{equation}
\label{n11}
 \langle n_{j\pm 1}\rangle = \rho \pm \rho_x + \tfrac{1}{2}\rho_{xx}+\cdots
 \end{equation}
and recast the set of difference-differential equations \eqref{DDE}
into a classical diffusion equation, namely Eq.~\eqref{DE} with
$D_1=1$. In higher dimensions, the cancellation still holds; in two
dimensions, for instance,
\begin{align}
\begin{split}
\frac{d \langle n_{i,j}\rangle}{dt} & =    \langle n_{i,j-1}\rangle  +
\langle n_{i,j+1}\rangle  + \langle n_{i-1,j}\rangle + \langle
n_{i+1,j}\rangle\\  & -    4\langle n_{i,j}\rangle\, .
\end{split}
\end{align}
Therefore the hydrodynamic description is again the classical
diffusion equation $\rho_t = \rho_{xx} + \rho_{yy}$. 

Consider now the simplest GEP different from the SEP, namely the GEP with $k=2$ on a one-dimensional lattice. 
The occupation number $n_j$ is either 0, or 1, or 2 when $k=2$. The
process $(n_j, n_{j+1})\Longrightarrow (n_j - 1, n_{j+1} + 1)$
proceeds with rate
\begin{equation}
\label{rate}
n_jF(n_{j+1}), \quad F(n) =1-\frac{n(n-1)}{2} .
\end{equation}
Therefore the average density evolves according to
\begin{align}
\begin{split}
\label{nj_av}
\frac{d \langle n_j\rangle}{dt} &=  \langle [n_{j-1}+ n_{j+1}]
F(n_j)\rangle  \\ &- \langle n_j
[F(n_{j-1})+F(n_{j+1})]\rangle\, .
\end{split}
\end{align}
In contrast with the case of the SEP, higher-order correlation
functions do not cancel as it is obvious from an explicit
representation of the right-hand side of \eqref{nj_av}:
\begin{align}
\begin{split}
\frac{d \langle n_j\rangle}{dt} &=  \langle n_{j-1}\rangle - 2\langle
n_j\rangle + \langle n_{j+1}\rangle \\ &+ \tfrac{1}{2}\langle
n_j  [n_{j-1}^2+n_{j+1}^2]-[n_{j-1}+ n_{j+1}] n_j^2\rangle\, .
\end{split}
\end{align}

It is often possible to advance for lattice gases of the \textit{gradient} type \cite{Spohn,KL99}. These are lattice gas models in which   
 the current  $J_{j,j+1}$ of particles moving from any site $j$ to $j+1$  can be written as a discrete gradient.
For instance, the SEP is the gradient lattice gas
since $ J_{j,j+1} = n_j-n_{j+1}$. 
For the GEP with $k=2$, the current 
\begin{equation}
J_{j,j+1}=n_j-n_{j+1} + \tfrac{1}{2}[n_{j+1}n_j^2-n_{j}n_{j+1}^2]
\end{equation}
is obviously not a discrete gradient. Generally all GEPs with $2\leq k<\infty$ are non-gradient lattice gases. 

We now outline the idea of a perturbative approach which we shall use to establish the diffusion coefficient in non-gradient lattice gases, and then return to the GEP. 

\subsection{Perturbative Approach}

For non-gradient lattice gases it is sometimes possible to study the hydrodynamic regime in the realm of a perturbative approach. The idea is to ignore correlations. This is true in the equilibrium. In the evolving state, the presence of local density gradients induces long-ranged correlations, but in numerous lattice gases these correlations vanish to first order in the density difference. This has been rigorously established (in all spatial dimensions) for lattice gases with hard-core exclusion \cite{Spohn83}, and it is expected to apply to a much larger class of models. There can be appreciable correlations in the earlier time regime, but we are interested in the hydrodynamic limit which, by definition, describes the evolution close to equilibrium. The vanishing of correlations in the first order in the density gradient implies that in the hydrodynamic regime our perturbative treatment leads to exact predictions for the diffusion coefficient. 

To appreciate the validity of a perturbative approach it is useful to compare the situation with kinetic theory \cite{Spohn,KRB10,fluid}. Recall that the Boltzmann equation, even though it is mean-field in nature (as it is based on the assumption of molecular chaos), is asymptotically exact in the hydrodynamic regime, so the emerging transport coefficients are exact. In the context of kinetic theory the main challenge is technical---even for dilute monoatomic gases  (e.g., for hard spheres gas), it has not been possible to \textit{extract} transport coefficients analytically \cite{fluid}. Lattice gases are much more tractable, so even for dense lattice gases the computing of the diffusion coefficient is occasionally feasible. The crucial ingredient is the understanding of the equilibrium state. We emphasize that for lattice gases of gradient type (e.g., for the Katz-Lebowitz-Spohn model with symmetric hopping \cite{KLS,Krug01} and for repulsion processes \cite{RP}) when computations using a Green-Kubo formula become feasible, the results for the diffusion coefficient \textit{agree} with predictions derived using the perturbative approach. Further, for lattice gases of non-gradient type whenever it was possible to apply the perturbative approach (see \cite{SOC:sing,EPA}), the predictions for the diffusion coefficient were again exact as it was evidenced through rigorous analyses, mappings to gradient type lattice gases, and comparisons with simulations.

\subsection{GEP with $k=2$}

To implement the perturbative approach for the GEP with $k=2$ on the one-dimensional lattice, we first replace \eqref{nj_av} by 
\begin{align}
\begin{split}
\label{nj_av_1d}
\frac{d \langle n_j\rangle}{dt}  & =  [\langle n_{j-1}\rangle + \langle
  n_{j+1}\rangle] \langle F(n_j)\rangle \\ & - \langle
n_j\rangle  [\langle F(n_{j-1}) \rangle +\langle F(n_{j+1})\rangle] .
\end{split}
\end{align}
In the hydrodynamic limit we write $\langle n_j(t)\rangle =\rho(x,t)$
and we use Eq.~\eqref{n11}  for $\langle n_{j\pm 1}(t)\rangle$ to
yield
\begin{equation}
\label{nn}
\langle n_{j-1}\rangle + \langle n_{j+1}\rangle = 2\rho+\rho_{xx} .
\end{equation}
Hereinafter,  we keep the terms which survive in the hydrodynamic limit,
e.g., in Eq.~\eqref{nn} we have dropped $\tfrac{1}{12}\rho_{xxxx}$ and
the following terms with higher derivatives. 

The average $\langle F(n)\rangle$ has a neat form
\begin{equation}
\label{F_av}
\langle F(n)\rangle = 1- P_2(\rho) ,
\end{equation}
which is obvious from the definition of the process (the hopping can
occur only when the target site hosts less than two particles). We shall
use the shorthand notation $1- P_2(\rho)=\Lambda_2(\rho)$.

In the hydrodynamic limit $\langle F(n_{j-1}) \rangle +\langle
F(n_{j+1})\rangle$ turns into
$\Lambda_2[\rho(x-1)]+\Lambda_2[\rho(x+1)]$, which is expanded to
yield
\begin{equation}
\label{FF_hydro}
 2 \Lambda_2(\rho)  + \Lambda_2'(\rho)\, \rho_{xx} +
 \Lambda_2''(\rho)\,\rho_x^2\, .
\end{equation}
Inserting all these expansions into \eqref{nj_av_1d} we arrive at
\begin{equation}
\label{rho_1d}
\rho_t = \left[\Lambda_2(\rho) - \rho\Lambda_2'(\rho)\right] \rho_{xx}
-\rho \Lambda_2''(\rho)\, \rho_x^2\, .
\end{equation}
This equation can be re-written as the diffusion equation \eqref{DE}
with diffusion coefficient
\begin{equation}
\label{D2}
D_2 = \Lambda_2(\rho) - \rho\Lambda_2'(\rho)\, .
\end{equation}
Recall that, for $k=2$, we have 
\begin{equation}
\label{rP2}
\rho = \frac{\lambda+\lambda^2}{1+\lambda+\tfrac{1}{2}\lambda^2}\,,
\quad \Lambda_2 = \frac{1+\lambda}{1+\lambda+\tfrac{1}{2}\lambda^2}
\end{equation}
from which we find an explicit expression for $\Lambda_2(\rho)$:
\begin{equation}
\label{L2}
\Lambda_2(\rho) =
\frac{ 1 - \rho + \sqrt{1+2\rho-\rho^2} }{2}\, .
\end{equation}
Inserting this into \eqref{D2} yields an explicit expression of the
diffusion coefficient
\begin{equation}
\label{D2_explicit}
D_2(\rho) =  \frac{1 + \rho +
  \sqrt{1+2\rho-\rho^2}}{2\sqrt{1+2\rho-\rho^2}}\, .
\end{equation}

We now consider the GEP with $k=2$ in arbitrary dimension.
In two dimensions, for instance, the average density satisfies
\begin{eqnarray*}
&&\frac{d \langle n_{i,j}\rangle}{dt}  =\langle ( n_{i-1, j} +
    n_{i+1,j} + n_{i,j-1}+ n_{i,j+1} ) F(n_{i,j})\rangle\\ 
&&-\langle  n_{i,j}  [F(n_{i-1, j})\! + \!F(n_{i+1,j})\! +\! F(n_{i,j-1})\!+\!
    F(n_{i,j+1})]\rangle
\end{eqnarray*}
which in the hydrodynamic limit becomes
\begin{equation}
\rho_t = \partial_x(D_2\,\rho_x)+\partial_y(D_2\,\rho_y)
\end{equation}
with $D_2$ given by Eq.~\eqref{D2} as in one
dimension. The same holds in any spatial dimension, namely the GEP is
described by the diffusion equation
\begin{equation}
\rho_t = \nabla\cdot (D_2 \nabla \rho),
\end{equation}
where the diffusion coefficient is given by a universal formula
\eqref{D2} valid in arbitrary dimension.  The symmetric GEP is
therefore \textit{isotropic} on the hydrodynamic scale; namely, it is
described by the scalar diffusion coefficient.

\subsection{GEP with arbitrary $k$}

For the GEP with arbitrary $k$ the analysis is similar to the one presented above. The process
$(n_j, n_{j+1})\Longrightarrow (n_j - 1, n_{j+1} + 1)$ proceeds with
rate \eqref{rate}, where we only need to modify $F(n)$ to 
\begin{equation}
\label{rate_k}
F(n) =1-\frac{n(n-1)\cdots (n-k+1)}{k!}\, .
\end{equation}

It suffices to consider the one-dimensional case as the results for
the diffusion coefficient are independent of the spatial dimensionality.
Equations \eqref{nj_av} and \eqref{nj_av_1d}, with $F(n)$ given by \eqref{rate_k}, remain valid.
Equations \eqref{nn}--\eqref{D2} also hold if we replace $\Lambda_2$
by $\Lambda_k$, the probability that a site is not fully occupied;  for instance,  Eq.~\eqref{F_av} becomes $\langle F(n)\rangle = 1-P_k(\rho)\equiv \Lambda_k(\rho)$. Thus the diffusion coefficient is indeed given by the announced expression \eqref{Dk}. For $k\geq 5 $ an
explicit expression for $\Lambda_k(\rho)$ is apparently impossible to
deduce, but we can use a   parametric expression \eqref{P_max} which
follows from \eqref{probs}, \eqref{density}, and the definition
$\Lambda_k(\rho)=1- P_k(\rho)$. 

\section{Stationary density profiles}
\label{sec:SDP}

In the previous section we calculated the diffusion coefficient for the GEP using a perturbative approach. 
In this section, we present a non-direct test of our predictions. Specifically, we shall calculate stationary density profiles in one and two dimensions and compare these theoretical predictions with simulation
results. We will show that the diffusion equation with the diffusion coefficient given by Eqs.~\eqref{Dk} and \eqref{P_max} provides an  accurate description of the system at a macroscopic scale.

\subsection{One-dimensional density profiles}

\begin{figure}
\centerline{ \includegraphics[width=85mm]{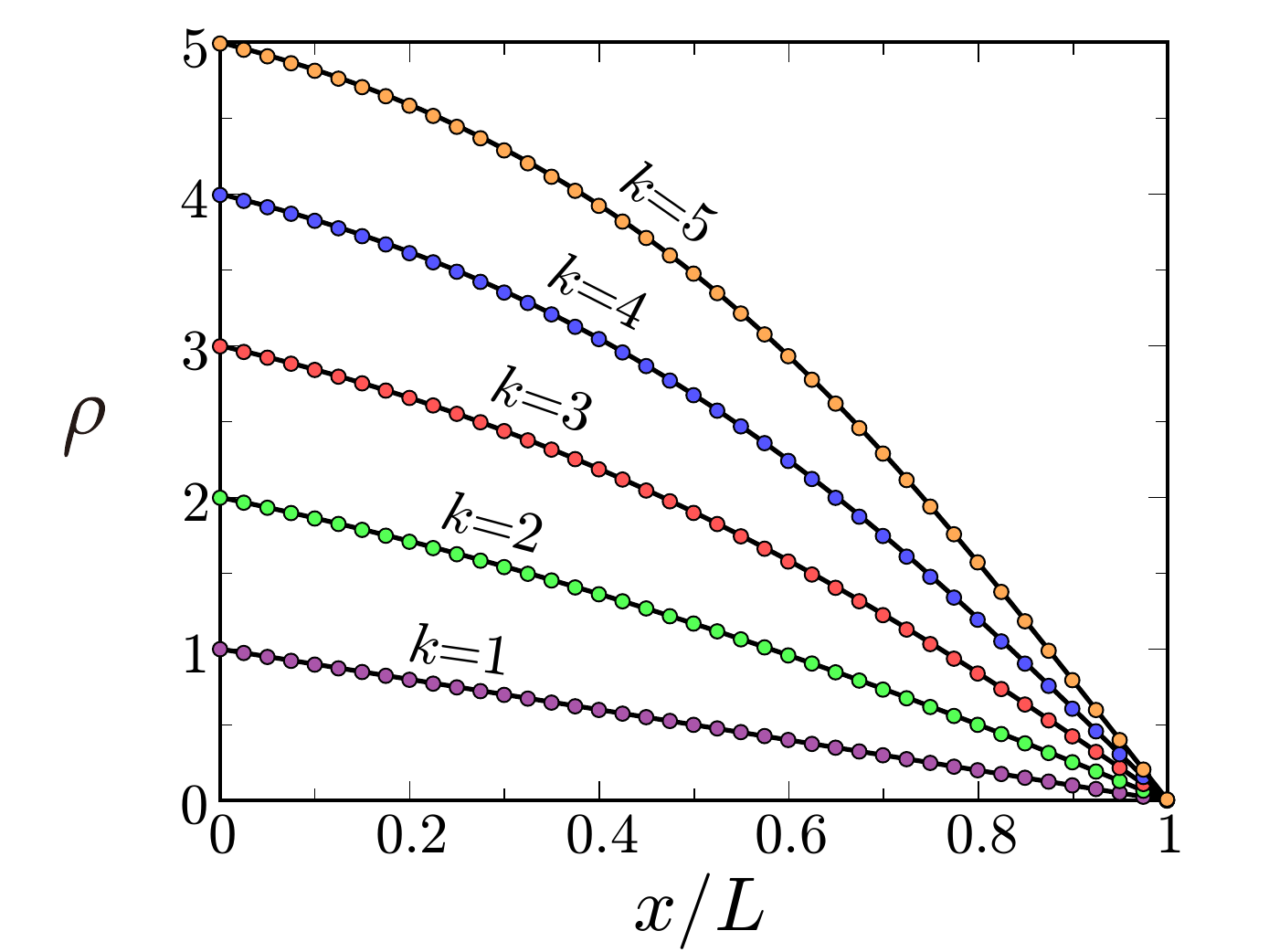} }
\caption{ Stationary density profiles versus $x/L$ for the GEP with
  $k=1,2,3,4,5$ on a segment with $L=10^3$.  The solid lines are
  theoretical predictions in the case of extremal boundary densities,
  $\rho_0=k$ and $\rho_1=0$. For the SEP ($k=1$), the density profile is linear; 
  for $k=2$, the density profile is given by \eqref{profile:k=2}, and generally it is extracted from
  Eq.~\eqref{Dk_profile_extreme}. Simulation results (shown by
  $\bullet$)  were obtained by averaging over the time window $
  5\times 10^6 \leq t\le 10^7$.}
\label{profile_fig}
\end{figure}

Consider the GEP on the interval $(0, L)$ with boundary conditions
\begin{equation}
\label{BC}
\rho(0) = \rho_0,  \quad \rho(L) = \rho_1 \, .
\end{equation}
Solving $D_k(\rho)\frac{d\rho}{dx}=\text{const}\,$,  subject to  \eqref{BC}, 
we obtain
\begin{equation}
\label{formal}
\frac{\int_{\rho_0}^\rho dr\,D_k(r)}{\int_{\rho_0}^{\rho_1}
  dr\,D_k(r)} = \frac{x}{L} \, .
\end{equation}
Let us focus on a special case when the right boundary is a  sink:
$\rho_1=0$.  To simplify formulas we write $\rho_0=n$. When
$k=2$, the integrals on the left-hand side of \eqref{formal} can be
explicitly determined to yield an implicit representation of the
stationary density profile $\rho(x)$:
\begin{equation*}
\frac{1 + \frac{\pi}{2} + \rho - \sqrt{1+2\rho-\rho^2} +
  2\arcsin\left(\frac{\rho-1}{\sqrt{2}}\right)} {1 + \frac{\pi}{2} + n
  - \sqrt{1+2n-n^2} + 2\arcsin\left(\frac{n-1}{\sqrt{2}}\right)}  =
1-\frac{x}{L}\, .
\end{equation*}
In the case of the maximal density on the left boundary, $\rho_0=n=2$,
we get
\begin{equation}
\label{profile:k=2}
\frac{1}{2} - \frac{\rho - \sqrt{1+2\rho-\rho^2} +
  2\arcsin\left(\frac{\rho-1}{\sqrt{2}}\right)} {\pi + 2} =
\frac{x}{L} \, ,
\end{equation}
see Fig.~\ref{profile_fig}. 

In the general case of arbitrary $k$ we use \eqref{Dk} and \eqref{P_max}
and establish the following parametric representation
\begin{equation}
\label{Dk_profile}
\frac{\int_0^\lambda d\mu\,[\Lambda_k(\mu)]^2} {\int_0^\ell
  d\mu\,[\Lambda_k(\mu)]^2}  = 1 - \frac{x}{L}\, , \quad n =
\ell\,\Lambda_k(\ell) \, .
\end{equation}
The maximal density on the left boundary, $\rho_0=n=k$, corresponds to
$\ell = \infty$. The density profiles \eqref{Dk_profile} in this
situation, 
\begin{equation}
\label{Dk_profile_extreme}
\frac{\int_0^\lambda d\mu\,[\Lambda_k(\mu)]^2} {\int_0^\infty
  d\mu\,[\Lambda_k(\mu)]^2}  = 1 - \frac{x}{L}\, , \quad \rho =
\lambda\,\Lambda_k(\lambda)
\end{equation}
are plotted	in Fig.~\ref{profile_fig}. 

\subsection{The GEP in an annulus}

For the GEP in the annulus $a\leq R\leq L$, we solve
$RD_k(\rho)\frac{d\rho}{dR}=\text{const}$,  subject to the boundary
conditions $\rho(a)=\rho_0$ and $\rho(L) = \rho_1$,  and get 
\begin{equation}
\label{formal_2d}
\frac{\int_{\rho_0}^\rho dr\,D_k(r)}{ \int_{\rho_0}^{\rho_1}
  dr\,D_k(r)} = \frac{\ln(R/a)}{\ln(L/a)} \,.
\end{equation}

\begin{figure}
\centerline{ \includegraphics[width=85mm]{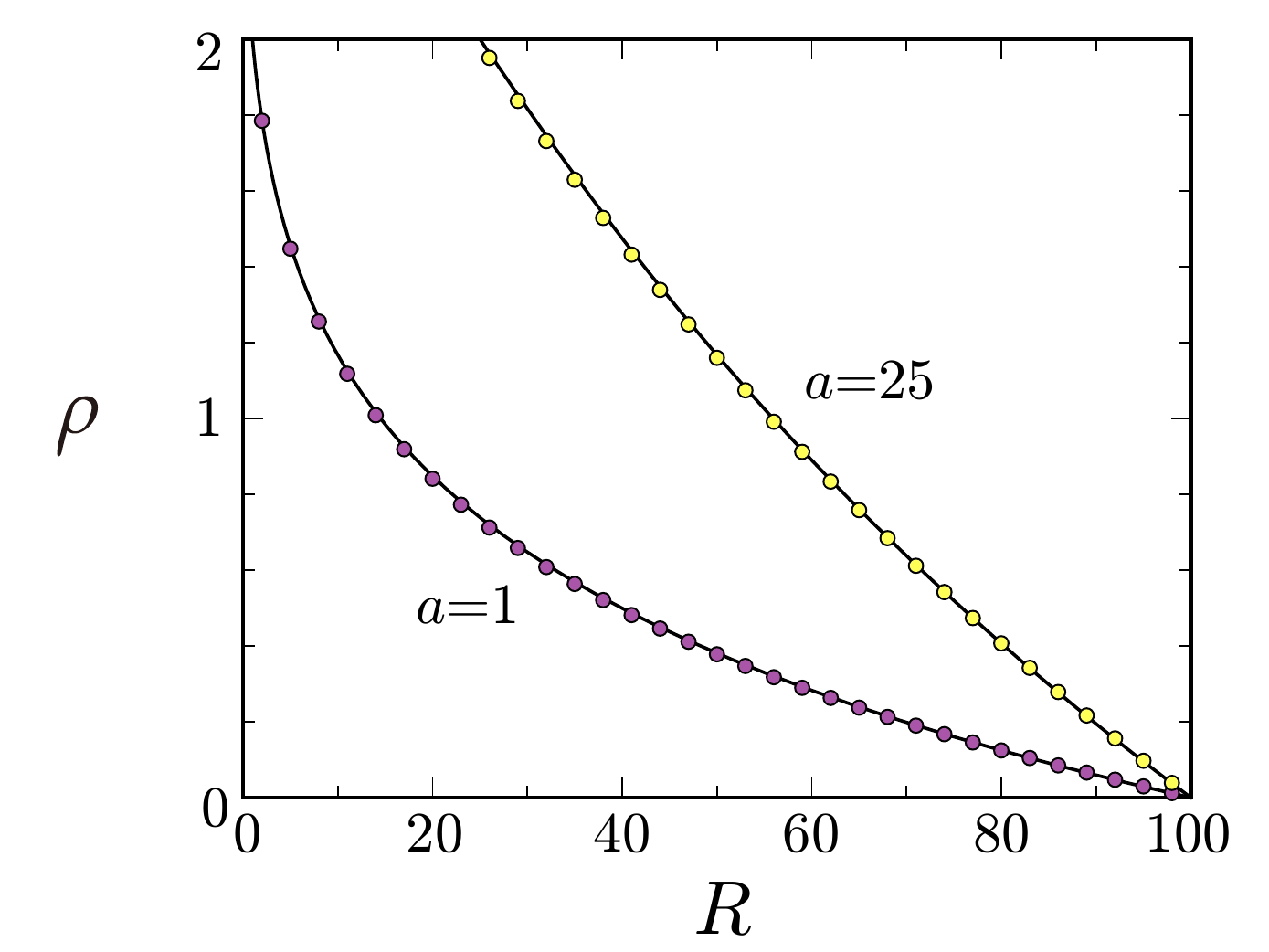} }
\caption{Stationary density profiles versus $R$ for the GEP with $k=2$
  in the annulus with external radius $L=100$.  Dots are simulation
  results which were obtained by averaging over $5\times 10^6\le t\le 
  10^7$. Solid lines are theoretical predictions given by
  \eqref{profile_2d:a=O(L)} with $\alpha=1/4$ in the case of $a=25$
  and by \eqref{profile_2d} when $a=1$.  }
\label{profile_D2_fig}
\end{figure}

Let us look more carefully at the case of $k=2$ with boundary
densities $\rho_1=0, \rho_0=2$ (the density on the inner circle is
maximal). We use dimensionless variables $\alpha = a/L$ and $\xi =
R/L$, so that  $0<\alpha\leq \xi\leq 1$. With these choices,
Eq.~\eqref{formal_2d} becomes 
\begin{equation}
\label{profile_2d:a=O(L)}
  \frac{1}{2} +
  \frac{\rho-\sqrt{1+2\rho-\rho^2}+2\arcsin\left(\frac{\rho-1}{\sqrt{2}}\right)}
       {\pi + 2}  =\frac{\ln \xi}{\ln \alpha }   \,.  
\end{equation}
This density profile is compared with simulation results on
Fig.~\ref{profile_D2_fig}.

For the GEP in the annulus $a\leq R\leq L$, the usage of the continuum
(diffusion equation) approach is somewhat questionable near the inner
circle if $a=O(1)$. Indeed, we cannot even talk about a circle on a
lattice if its radius is comparable with the lattice
spacing. Nevertheless, let us use Eq.~\eqref{formal_2d}, again with
$k=2$ and $(\rho_0,\rho_1)=(2,0)$, in the extreme case of
$a=1$. Equation \eqref{formal_2d} becomes 
\begin{equation}
\label{profile_2d}
\frac{1}{2} -  \frac{\rho - \sqrt{1+2\rho-\rho^2} +
  2\arcsin\left(\frac{\rho-1}{\sqrt{2}}\right)} {\pi + 2}  =\frac{\ln
  R }{\ln L }   \,.
\end{equation}
Choosing the inner radius equal to lattice spacing is essentially
equivalent to the simplest lattice setting with reservoir connected to
the origin and postulating that whenever a particle leaves the origin,
a particle from reservoir is immediately added, so the density at the
origin remains maximal $\rho_0=2$. There is also a sink at the circle
$R=L$; that is, whenever a particle at a site on distance $<L$ hops
and gets outside this circle, it leaves the system forever. On
distances $R\gg 1$ the profile \eqref{profile_2d} should become
asymptotically exact. Figure \ref{profile_D2_fig} shows an excellent
agreement between theory and simulations over the entire range $1\leq
R\leq L$.

To emphasize the difference between one and two dimensions let us
consider the GEP with $k=2$ and boundary densities
$(\rho_0,\rho_1)=(2,0)$ and compare the density profiles
\eqref{profile:k=2} and \eqref{profile_2d}. In one dimension, the
intermediate density $\rho_* = (\rho_0+\rho_1)/2$, i.e., $\rho_* = 1$
in our case, is reached at 
\begin{equation}
\frac{x_*}{L} = \frac{1}{2} + \frac{\sqrt{2}-1}{\pi + 2} =
0.580561\ldots ,
\end{equation}
while in two dimensions this happens at
\begin{equation}
\frac{\ln R_*}{\ln L} = \frac{1}{2} + \frac{\sqrt{2}-1}{\pi + 2} ,
\end{equation}
which is much closer to the source, $R_*\sim L^{0.580561}$. 

Second, we compare the total (average) number of particles. In one
dimension we integrate by part to get
\begin{equation}
N=\int_0^L dx\,\rho(x)= \int_0^2 d\rho\, x(\rho) .
\end{equation}
Using \eqref{profile:k=2} we perform the integration and find 
\begin{equation}
N = \frac{3\pi + 2}{2\pi+4}\,L .
\end{equation}
In two dimensions, we similarly find
\begin{equation}
\label{N_2d}
N=\int_0^L dR\,2\pi R \rho(R)= \pi\int_0^2 d\rho\, R^2(\rho)  \,.
\end{equation}
The dominant part of the integral in \eqref{N_2d} is gathered near
$\rho=0$. Expanding the left-hand side of \eqref{profile_2d} we find 
\begin{equation}
\frac{\ln R}{\ln L} = 1 - \frac{2}{\pi + 2}\,\rho - \frac{1}{3(\pi +
  2)}\,\rho^3 + \frac{1}{2(\pi + 2)}\,\rho^4+\cdots  \,.
\end{equation}
Equation \eqref{N_2d} becomes
\begin{equation}
\frac{N}{\pi L^2}\simeq \int_0^\infty d\rho\,  \exp\!
\left[-   \ln L\, \frac{4\rho +\tfrac{2}{3}\rho^3   - \rho^4   }{\pi + 2}\right] ,
\end{equation}
which gives 
\begin{equation}
\label{N_2d_asymp}
N =\frac{\pi(\pi + 2)}{4}\,\frac{L^2}{\ln L}
\left[1 + \frac{C_2}{(\ln L)^2}+\frac{C_3}{(\ln L)^3}+\cdots\right]  
\end{equation}
with $C_2=-\left(\frac{\pi+2}{4}\right)^3,
C_3=6\left(\frac{\pi+2}{4}\right)^4$, etc. Thus the convergence to the
leading asymptotic behavior is slow in two dimensions. 

For arbitrary $k$, let us choose again $\rho_1=0$ and $\rho_0=k$. the
density profile is implicitly given by 
\begin{equation}
\frac{\int_0^\lambda d\mu\,[\Lambda_k(\mu)]^2} {\int_0^\infty
  d\mu\,[\Lambda_k(\mu)]^2} = 1-\frac{\ln R}{\ln L}  \,.
\end{equation}
In the small $\rho$ limit, we get
\begin{equation}
\frac{\ln R}{\ln L} = 1 - \frac{\rho}{I_k}+\cdots, \quad
I_k=\int_0^\infty d\mu\,[\Lambda_k(\mu)]^2  \,,
\end{equation}
and the leading asymptotic behavior of the total average number of
particles is
\begin{equation}
N \simeq \frac{\pi I_k}{2}\,\frac{L^2}{\ln L}
\end{equation}
where the coefficients $I_k$ can be evaluated numerically
(e.g. $I_3=4.29139\ldots$). 

\section{Self-Diffusion Coefficient}
\label{sec:SDC}

Even an equilibrium situation (in which the density is spatially uniform) possesses interesting non-equilibrium features. One important example is the phenomenon of self-diffusion. In this section we investigate the evolution of a tagged particle in the GEP at equilibrium.  We assume that the tagged particle is identical to the host particles, so it merely carries a tag. Asymptotically, the tagged particle exhibits a diffusive behavior, so it suffices to compute the coefficient of
self-diffusion. This problem is easy to pose, but there has been little progress even for simplest lattice gases. For instance, the coefficient of self-diffusion is unknown for the SEP in two and higher dimensions, it is only known \cite{LOV_01} that the coefficient of self-diffusion is a smooth function of the density.

Consider first the one-dimensional case. We tag a particle which is initially at $x(0)=0$
(without loss of generality) and we look at its position $x(t)$ in the
long time limit.
Generically, we expect a diffusive behavior.
Thus the first two averages are $\langle x\rangle = 0$ and $\langle x^2\rangle\sim t$,
 and it suffices to determine the self-diffusion coefficient
\begin{equation}
\label{SD:def}
\lim_{t\to\infty}\frac{\langle x^2\rangle}{2t} \, = \,
\mathcal{D}_k(\rho).
\end{equation}
The self-diffusion coefficient  $\mathcal{D}_k$ generally differs from
the diffusion coefficient $D_k$. We have $\mathcal{D}_\infty =
D_\infty=1$ for non-interacting random walks. For $k<\infty$, the
inequality $\mathcal{D}_k<D_k$ is physically apparent, although it may
be difficult to prove. 

For the SEP in one dimension,  the self-diffusion coefficient vanishes:
$\mathcal{D}_1=0$. Indeed,  the ordering between the particles
is conserved and this leads to anomalously slow
sub-diffusive behavior \cite{Harris_65,Levitt_73,PMR_77,AP_78,Arr_83}:
$\langle x^2\rangle_{\text{SEP}, \,\, d=1}\sim t^{1/2}$.
This is an exceptional feature; the normal diffusion is recovered for the SEP
in dimensions higher than 1. For the GEP with $k\geq 2$, the phenomenon of self-diffusion is not
pathological even in one dimension, viz. the self-diffusion
coefficient $\mathcal{D}_k(\rho)$ is positive. Moreover,
$\mathcal{D}_k(\rho)$ is a monotonically decreasing function of $\rho$
in the interval $0<\rho<k$ with asymptotic behaviors
\begin{equation}
\label{SD_asymp}
\mathcal{D}_k(\rho) = 
\begin{cases}
1 & \rho\to 0 ,\\ 
0 & \rho\to k .
\end{cases}
\end{equation}

For lattice gases in $d>1$ dimensions, the spread of the tagged particle is generically described by a matrix. For the GEP on the hyper-cubic lattice, and generally for lattice gases on $\mathbb{Z}^d$ where each particle occupies only one site, one expects the self-diffusion process to be isotropic on the hydrodynamic scales. This has been proved only for the SEP, see \cite{KV_86}, so it remains conjectural for the GEPs with $2\leq k<\infty$. 

\begin{figure}
\centerline{ \includegraphics[width=85mm]{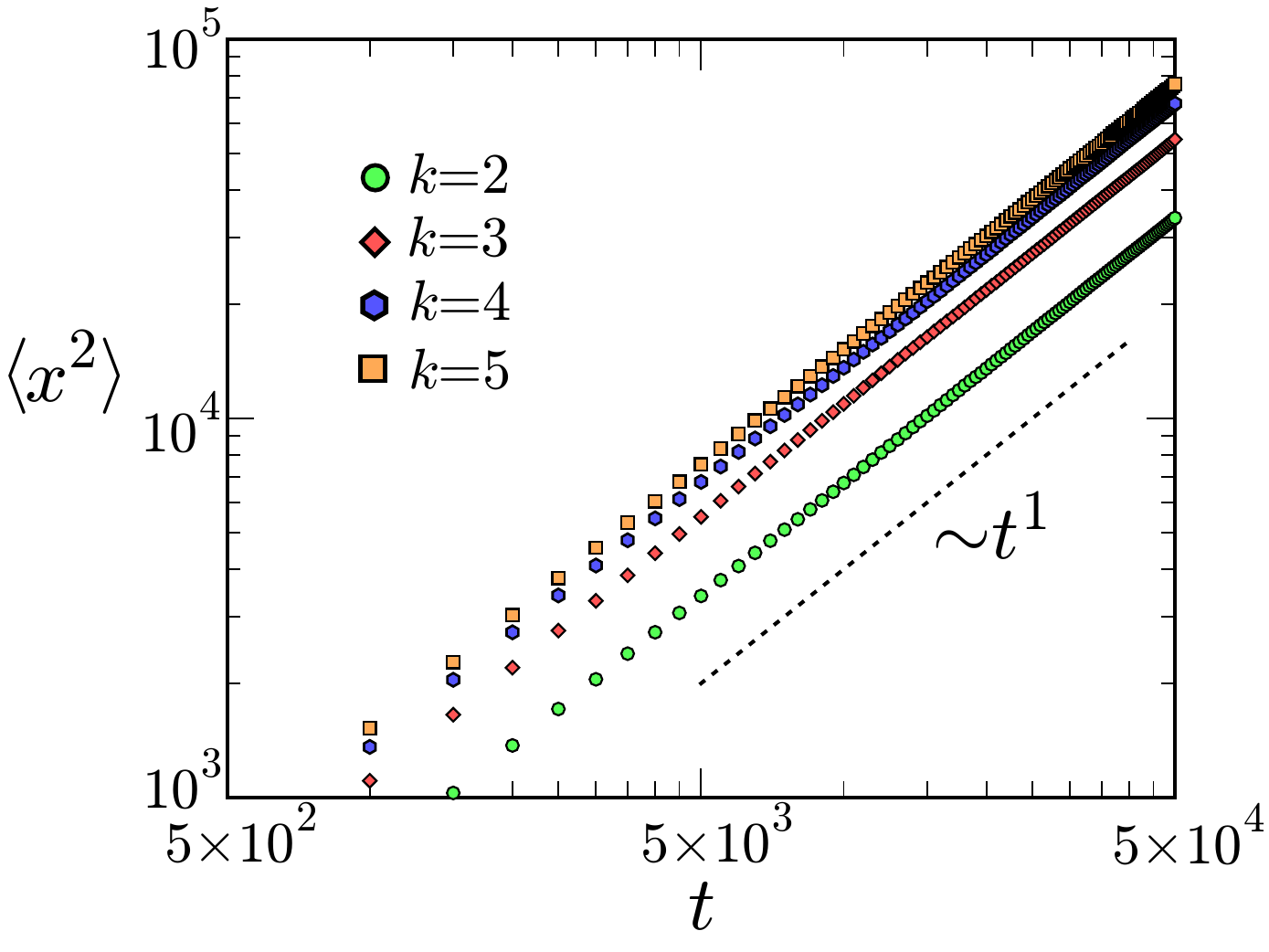}}
\caption{Mean-square displacement of the tagged particle vs
  time for the GEP in one dimension.  The maximal occupancy varies
  between $k=2$ and $k=5$.  Simulations were performed on the ring of
  length $10^3$.}
\label{GEP-Xt2_fig}
\end{figure}

There is one important feature which distinguishes the self-diffusion coefficient from the diffusion coefficient, viz. the self-diffusion coefficient certainly depends on the dimensionality: $\mathcal{D}_k(\rho, d)$. The extreme density behaviors of the self-diffusion coefficient $\mathcal{D}_k(\rho, d)$ with $k\geq 2$ are universal and given by \eqref{SD_asymp} in all dimensions.

\begin{figure}
\centerline{ \includegraphics[width=88mm]{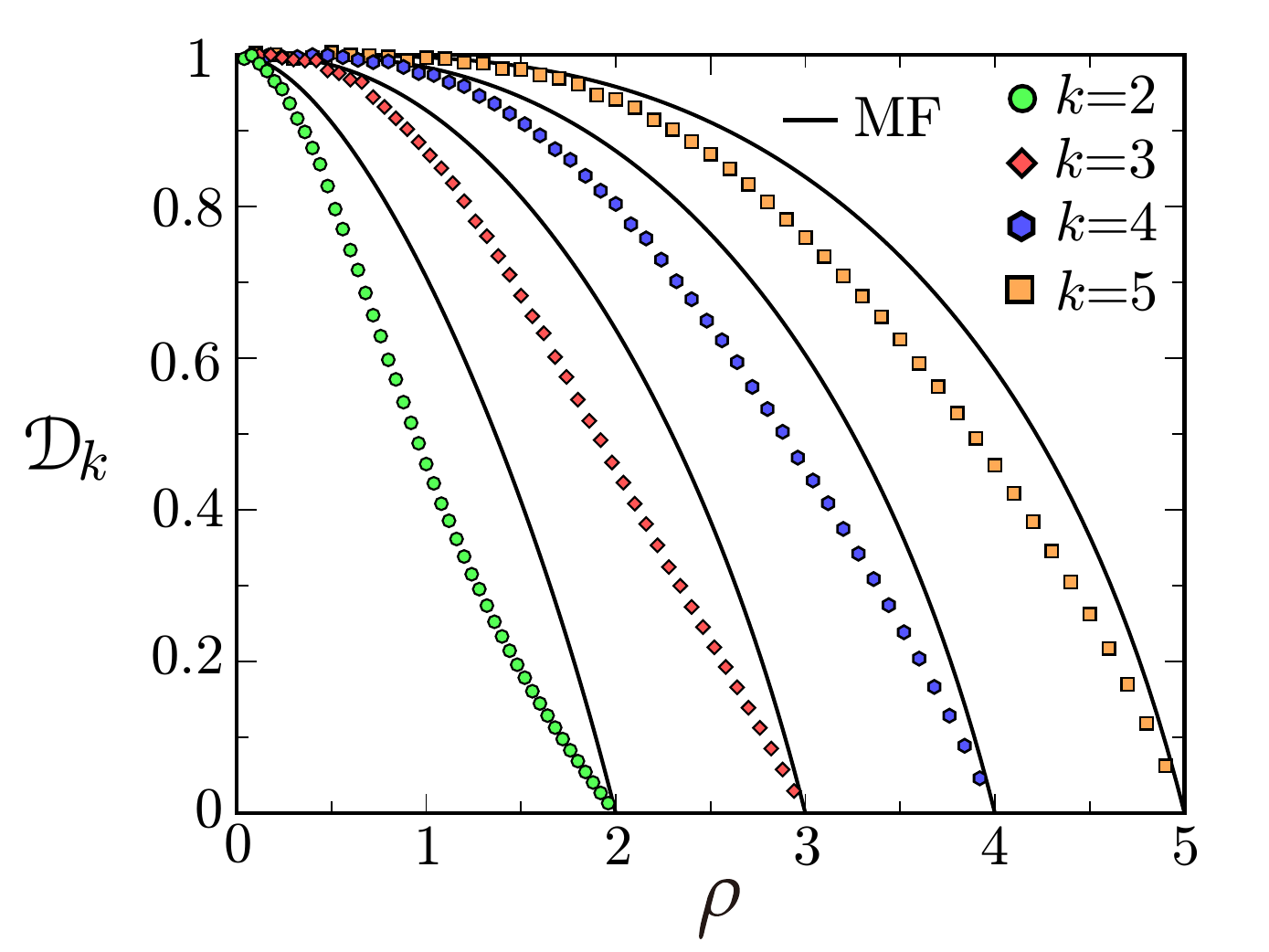}}
\caption{Coefficient of self-diffusion vs density for the GEP in one dimension;
  the maximal occupancy varies between $k=2$ and $k=5$.  Dots
  represent simulation results for the GEP  in one dimension. Solid
  lines are the mean-field predictions,   Eq.~\eqref{SD:MF}.}
\label{GEP-self-1d_fig}
\end{figure}

\begin{figure}
\centerline{ \includegraphics[width=88mm]{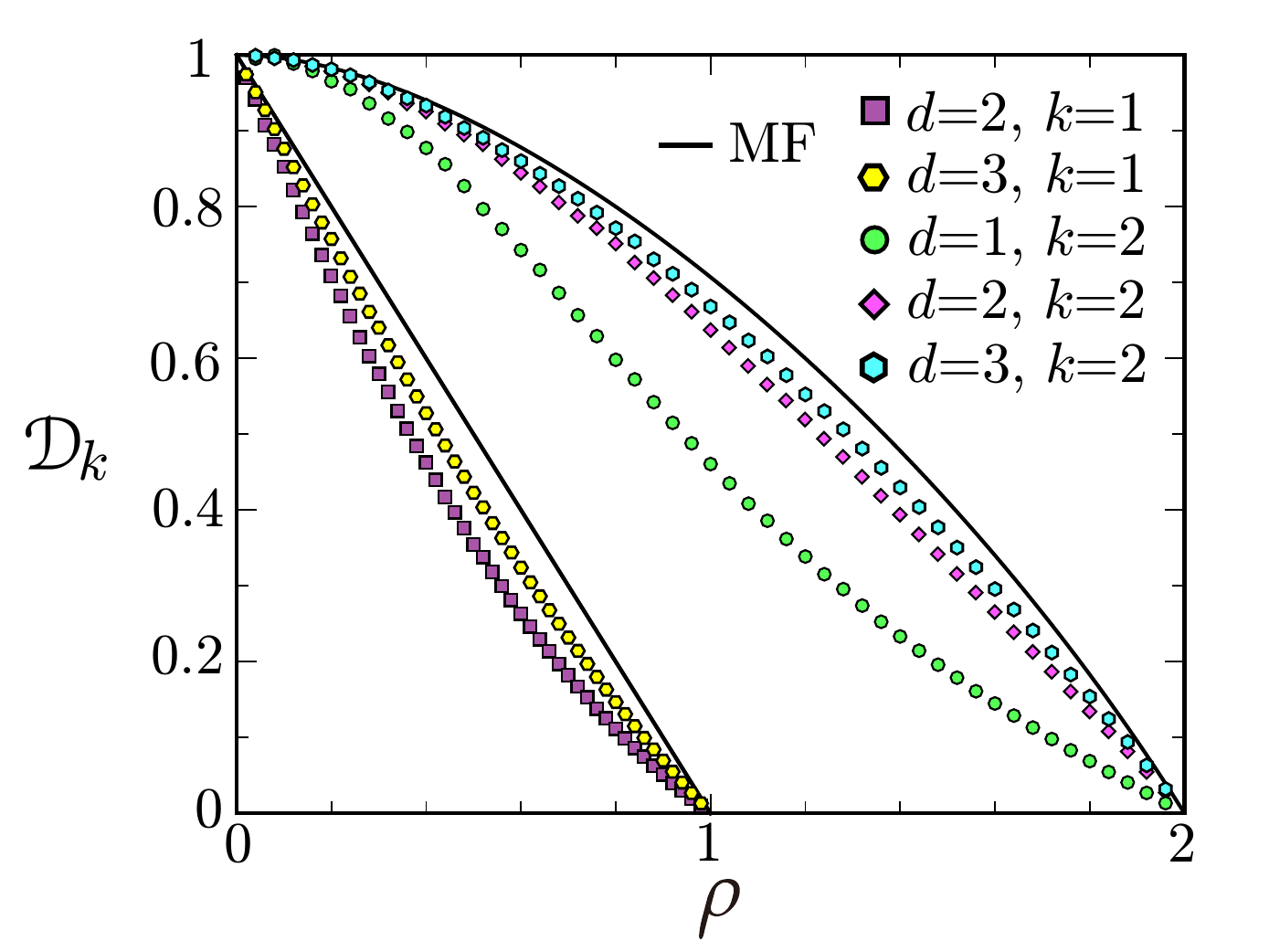}}
\caption{Coefficient of self-diffusion vs density for the SEP in two and three dimensions, and for the GEP 
  with $k=2$ in dimensions $d=1, 2, 3$.  Dots represent simulation results. Solid
  lines are the mean-field predictions: $\mathcal{D}_1^\text{MF} = 1 - \rho$ for the SEP,
  while for the GEP with $k=2$ the mean-field prediction 
  $\mathcal{D}_2^\text{MF} = \Lambda_2(\rho)$ acquires the explicit form
  \eqref{L2}.}
\label{SD-123d_fig}
\end{figure}

We performed simulations to probe the self-diffusion coefficient on lattices with $L^d$ sites, in $d=1,2,3$ dimensions, with periodic boundaries.
The sizes of the simulated systems are $L=10^3,~32,~10$  for $d=1,2,3$, respectively.  
The number of simulation runs for each set of parameters $ (k,d,\rho)$ is  $2\times 10^5/(L^d \rho)$, and we tagged all the $L^d \rho$ particles in each run. Thus $\langle \cdot \rangle $ is the average over effectively $2\times 10^5$ tagged particles. We  checked the validity of the diffusive scaling up to $t=5\times 10^4$, as shown in Fig.~\ref{GEP-Xt2_fig} for $\rho = 3k/5$  and $d=1$.
(As long as $t\ll L^{2} = 10^6$, finite size effects can be safely ignored.) We calculated 
$\frac{\langle \mathbf{r}^2 \rangle}{2d\cdot t}$ by using data in the $0\le t\le 5\times 10^4$ time window.
The results are shown in Figs.~\ref{GEP-self-1d_fig} and \ref{SD-123d_fig}. 
 
As a reference point, it is useful to have a mean-field prediction. To
derive the mean-field prediction for the self-diffusion coefficient of
the GEP with an arbitrary maximal occupancy $k$ we recall that a site is occupied
by $k$ particles with probability $P_k$, so it can be a destination
site with probability $1-P_k$. Therefore for a tagged particle, the
hopping rate to each neighboring site appears to be $1-P_k(\rho)\equiv \Lambda_k(\rho)$, which
tells us that the self-diffusion coefficient  is
\begin{equation}
\label{SD:MF}
\mathcal{D}_k^\text{MF} = \Lambda_k =
\frac{ Z_{k-1}(\lambda) }{ Z_k(\lambda)} =
\frac{\rho}{\lambda}\,.
\end{equation}
In particular, $\mathcal{D}_1^\text{MF}=1-\rho$ for the SEP, while for $k=2$ the mean-field prediction 
$\mathcal{D}_2^\text{MF}=\Lambda_2(\rho)$ is given by the explicit formula \eqref{L2}. For non-interacting random walks Eq.~\eqref{SD:MF} yields $\mathcal{D}_\infty^\text{MF}=1$, and this is the only case when the prediction is exact. In all other cases ($1\leq k<\infty$), the prediction of Eq.~\eqref{SD:MF} is \textit{not} exact. 

To appreciate the mean-field nature of the prediction \eqref{SD:MF} we first note that for every
site, the probability that any neighboring site contains less than $k$
particles is indeed $1-P_k$, and these probabilities are
uncorrelated. So if we pick a particle and mark it with a tag, it
appears that this particle is indeed diffusing with the coefficient
equal to $1-P_k$.  But we must keep the identity of the tagged
particle. This already causes the problem---immediately after the
tagged particle has undergone the first jump, the site from which it
jumped will be surely occupied by less than $k$ particles. In the
$d\to\infty$ limit this is irrelevant, but for any finite dimension
the derivation of  Eq.~\eqref{SD:MF} involves an uncontrolled
approximation. We thus realize that  Eq.~\eqref{SD:MF}  only provides a
mean-field approximation. To summarize, the prediction \eqref{SD:MF}
satisfies the following  properties:
\begin{enumerate}
\item It agrees with the limiting behaviors \eqref{SD_asymp}.
\item It appears to be an upper bound for all $d\geq 1$.
\item It becomes exact in the $d\to\infty$ limit.  This justifies
  calling \eqref{SD:MF} a mean-field prediction. 
\item It is also exact for non-interacting random walks ($k=\infty$). 
\end{enumerate}

The validity of the first property easily follows from
Eq.~\eqref{SD:MF}, and it is also seen (in the one-dimensional case)
from Fig.~\ref{GEP-self-1d_fig}. The second property seems very
plausible, but  has not been proved; it is supported by simulation results; see
Figs.~\ref{GEP-self-1d_fig} and \ref{SD-123d_fig}. The validity of the third and fourth
properties is obvious. 

Figures ~\ref{GEP-self-1d_fig} and \ref{SD-123d_fig} indicate that the disagreement
between the actual behaviors and the mean-field predictions is most
pronounced in one dimension, so the mean-field estimate \eqref{SD:MF}
provides a good approximation in two and three dimensions.

Assuming that the mean-field estimate \eqref{SD:MF} provides
qualitatively correct small and large $\rho$ behaviors also in finite
dimensions, we anticipate that
\begin{equation}
\lim_{\rho\downarrow 0} \frac{1-\mathcal{D}_k(\rho, d)}{\rho^k} =
A_k(d), \ \  \lim_{\rho\uparrow k} \frac{\mathcal{D}_k(\rho,
  d)}{k-\rho} = B_k(d)\,.
\end{equation}

Little is known about these amplitudes. For $d=\infty$,
\begin{equation}
A_k(\infty) = \frac{1}{k!}\,, \  \ B_k(\infty) = 1\,,
\end{equation}
because the mean-field prediction becomes exact when $d=\infty$. Since the mean-field estimate \eqref{SD:MF} apparently provides an upper bound, we expect that  $A_k(d)>\frac{1}{k!}$ and $B_k(d)<1$.

\section{Summary}
\label{sec:Sum}

We investigated generalized exclusion processes with symmetric
nearest-neighbor hopping parametrized by an integer $k$, the maximal
occupancy. Specifically, we studied a class of such processes
interpolating between the symmetric exclusion process ($k=1$) and
non-interacting random walkers ($k=\infty$).  For these lattice gases
the hydrodynamic behavior is governed by a diffusion equation. We
computed the diffusion coefficient $D_k$ and showed that for every $k$, it
does not depend on the spatial dimension, but it does depend on $k$. We showed that, 
apart from the extreme cases of $k=1$ and $k=\infty$, the diffusion coefficient depends on the density. 

We studied numerically the self-diffusion phenomenon for the GEPs in one, two, and three dimensions.  An interesting challenge is to compute the self-diffusion coefficient for the GEPs with $k\geq 2$ in one dimension. In two and higher dimensions this problem seems intractable, even for the SEP in two dimensions the coefficient of self-diffusion is unknown; it has only been established that $\mathcal{D}_1(\rho, d)$ is a smooth function of the density \cite{LOV_01}. In the one-dimensional setting, the tagged particle in the case of the SEP undergoes an anomalously slow sub-diffusive behavior which is well understood, and even large deviations of the displacement of the tagged particle have been probed (see e.g. \cite{SV_13,KMS_14} and references therein). Therefore there is a hope that the self-diffusion phenomenon in the one-dimensional GEP is also analytically tractable. 

In addition to the diffusion coefficient, a second transport coefficient, the mobility (or conductance)  $\sigma(\rho)$, plays an important role in the macroscopic fluctuation theory \cite{MFTReview}. The knowledge of $\sigma(\rho)$ is required if one wants to understand fluctuations around the (deterministic) hydrodynamic behaviors, including large deviations. We leave the determination of the mobility $\sigma(\rho)$  for future studies of fluctuations and large deviations in the GEPs.

In this article we considered only the GEP with symmetric hopping. One would like to understand the asymmetric version of the GEP. The problem is that in the driven case the structure of the steady states is unknown: it is not a product measure anymore, even on a ring \cite{Timo}. One possibility is to modify the rules of the GEP to make the structure of the steady states more accessible. An interesting variant is to employ a drop-push dynamics when a hopping attempt to a fully occupied neighboring site is not rejected, but instead the particle proceeds in the same direction and lands at the closest site which is not fully occupied \cite{Barma_drop,Barma}; a similar process was suggested, and studied for $k=1$, in the context of self-organized criticality \cite{SOC:sing}. For these GEPs with drop-push dynamics the structure of the steady states is known in the general case of asymmetric hopping \cite{Barma_drop,Barma}. The symmetric version of the GEPs with drop-push dynamics also deserves further analysis, e.g., one would like to compute the diffusion coefficient for this class of lattice gases. 

\medskip
\textbf{Acknowledgments.}\quad 
  One of the authors (CA) thanks  Chihiro Matsui  for a stimulating discussion. KM and PLK are grateful to the Galileo Galilei Institute for Theoretical Physics for the hospitality and the INFN for partial support during the completion of this work. The work of P.L.K. was partly supported by grant No.~2012145
from the US--Israel Binational Science Foundation (BSF).


\begin{thebibliography}{99}
\bibitem{critical}
  More precisely, $D(\rho)>0$ away from a critical density where the equilibrium lattice gas undergoes 
  a phase transition. Such a phase transition may occur in $d\geq 2$ dimensions; e.g., it occurs in exclusion 
  processes with infinitely strong repulsion between particles in adjacent sites. 
  However, there are no phase transitions in GEPs which we study. In these lattice gases 
  a stronger bound, $D_k(\rho)\geq 1$,  is actually valid for all $k$ and $\rho$.
\end{thebibliography}
\end{document}